\documentclass[letter]{aa} 

\usepackage{graphicx}
\usepackage{txfonts}
\usepackage{hyperref}
\usepackage{isotope}

\newcommand{\msun}{$\rm M_{\odot}$}

\newcommand{\stocl}{\isotope[34]{S}($p$,$\gamma$)\isotope[35]{Cl}}
\newcommand{\stocll}{\isotope[32]{S}($p$,$\gamma$)\isotope[33]{Cl}}
\newcommand{\ptosi}{\isotope[31]{P}($p$,$\alpha$)\isotope[28]{Si}}
\newcommand{\artok}{\isotope[38]{Ar}($p$,$\gamma$)\isotope[39]{K}}

\title{The SPAr burning: proton captures powering carbon--oxygen shell mergers in massive stars}
\author{Lorenzo Roberti}
\date{May 2025}

\begin{document}

\author{
L. Roberti\inst{1,2,3,4,5} 
\and M. Pignatari\inst{2,3,6,5}
}

\institute{
Istituto Nazionale di Fisica Nucleare - Laboratori Nazionali del Sud, Via Santa Sofia 62, Catania, I-95123, Italy
\and Konkoly Observatory, Research Centre for Astronomy and Earth Sciences, HUN-REN, Konkoly Thege Miklós út 15-17, Budapest, H-1121, Hungary 
\and CSFK HUN-REN, MTA Centre of Excellence, Konkoly Thege Miklós út 15-17, Budapest, H-1121, Hungary
\and Istituto Nazionale di Astrofisica – Osservatorio Astronomico di Roma, Via Frascati 33, Monte Porzio Catone, I-00040, Italy
\and NuGrid Collaboration, \url{http://nugridstars.org}
\and University of Bayreuth, BGI, Universitätsstraße 30, 95447 Bayreuth, Germany
}

\date{}
\date{Received August 22, 2025; accepted November 4, 2025}

\abstract
{Carbon-oxygen (C--O) shell mergers in massive stars play a crucial role in both nucleosynthesis and the final stages of stellar evolution. These convective-reactive events significantly alter the internal structure of the star shortly before core collapse.}  
{We investigate how the enhanced production of light particles (especially protons) during a C--O shell merger, relative to classical oxygen shell burning, affects the energy balance and evolution of the convective shell.}  
{We derive the budget for direct and reverse nucleosynthesis flows across all relevant nuclear reactions from stellar evolution models, and we assess the relative energy produced.}
{We find that proton capture reactions on \isotope[32,34]{S}, \isotope[31]{P}, and \isotope[38]{Ar} (SPAr) dominate the nuclear energy production in typical C--O shell mergers as predicted by 1D stellar models. Their combined energy output is approximately 400 times greater than that of C and O fusion under the same conditions.}  
{Our results highlight the critical importance of including these proton-capture reactions in simulations of convective-reactive burning. This work suggests that excluding their contribution can lead to inaccurate modeling of the dynamics and nucleosynthesis in advanced stellar evolutionary phases. Such results will need to be confirmed by new 1D stellar simulations and 3D hydrodynamics models. }

\keywords{stars: massive -- stars: evolution -- stars: interiors -- stars: nucleosynthesis -- Galaxy: abundances -- supernovae: general}

\maketitle

\section{Introduction} \label{sec:intro}

    The advanced stages of the evolution of a massive star ($\rm M>8-9$ \msun) are characterized by a complex interplay between nuclear burning, neutrino cooling, and formation of convective zones. Stars less massive than 25--30 \msun\ can often develop efficient O burning shells that penetrate the outer C-- and Ne--rich layers and eventually merge in a single extended mixed zone a few days or even hours before the core-collapse supernova \citep[CCSN, e.g.,][]{rauscher:02,ritter:18,roberti:24,roberti:25}. This occurrence, known often as C--O shell merger, has important implications both on the nucleosynthesis and on the structure of the star before and during the CCSN stage. 

    The signature of a C--O shell merger in the chemical composition of the CCSN ejecta is represented by an enhanced production of the elements synthesized by the O-burning (Si, S, Ar, Ca), as well as some odd-Z elements \citep[P, Cl, K, Sc,][]{ritter:18,roberti:25}. Radioactive nuclei as \isotope[40]{K} can be efficiently produced as well \citep[][and references therein]{issa:25b}. Eventually, due to the efficient activation of photodisintegrations at the bottom of the O shell ($\rm T\sim 2.5-2.8\ GK$), the "cold" $\gamma-$process can occur and produce the $p$--nuclei with $\rm A>110$ \citep{ritter:18,roberti:23,issa:25a}. Moreover, this production is mostly preserved by the CCSN shock, due to the extended radius of the newly formed mixed region \citep{roberti:24b}.
    
    Recent works \citep{andrassy:20,yadav:20,rizzuti:22,rizzuti:23,rizzuti:24} explored the occurrence of C--O shell mergers in 3D hydrodynamical models, using as an initial condition the structure of the O, Ne, and C shells obtained by a 1D model, remapped on a three-dimensional grid. The evolution of the shell interaction is then often followed with the adoption of a minimal nuclear network including only the main nuclear reactions involving the isotopes \isotope[12]{C}, \isotope[20]{Ne}, and \isotope[16]{O} \citep[see, e.g.,][]{andrassy:20,rizzuti:22}, with the aid of additional boosting factors for the energy generation to accelerate the simulation. \cite{rizzuti:23,rizzuti:24} instead employed for the first time an explicit 12-isotope nuclear network (n, p, \isotope[4]{He}, \isotope[12]{C}, \isotope[16]{O}, \isotope[20]{Ne}, \isotope[23]{Na}, \isotope[24]{Mg}, \isotope[28]{Si}, \isotope[31]{P}, \isotope[32]{S}, and \isotope[56]{Ni}), including all the nuclear reactions linking these isotopes. All these works found quantitative differences with 1D models, especially in the much higher convective velocity achieved in 3D models and the natural entrainment of surrounding material from the edges of the convective zones, eventually facilitating the merger to occur \citep[see also][]{jones:17}. 

    Furthermore, this merger may introduce asymmetries in the stellar structure and lead to a significant density drop at the Si–O interface, potentially favoring the onset of the CCSN explosion\citep[see, e.g.,][]{couch:13,wang:22,boccioli:23,laplace:25}. 
    
    In a C--O shell merger, the simultaneous activation of \isotope[12]{C}+\isotope[12]{C}, \isotope[12]{C}+\isotope[16]{O}, \isotope[20]{Ne}($\gamma$,$\alpha$)\isotope[16]{O}, and \isotope[16]{O}+\isotope[16]{O}, leads to a large production of the light particles $p$, $\alpha$ and $n$. Probabilities of charged particle reactions drastically increase compared to the heavy ion fusion reactions, that have to overcome a larger Coulomb barrier to occur. These reactions not only influence the nucleosynthesis as described above. In this work we will show that some proton-capture reactions can significantly alter the nuclear energy release in the shell merger, even overpowering the energy generated by C and O fusions.

\section{Astrophysical conditions} \label{sec:model}

    We explore the astrophysical conditions of O shell burning before, during, and after the occurrence of a C--O shell merger. We take as an example the 15 \msun\ stellar model at extremely low metallicity ($\rm Z_{ini}\simeq10^{-5}Z_{\odot}$), with an initial equatorial rotation velocity of $\rm v_{ini}=600\ km\ s^{-1}$ from \cite{roberti:24} (see Appendix \ref{app:1} for more details). 
    
    \figurename~\ref{fig:abu} shows the chemical composition inside the O burning shell at three different times before the core collapse: during the classical O burning in shell (i.e., without any ingestion of Ne or C rich material, $\rm \Delta t_{coll} = 3.47\times10^{5}\ s$, upper panel); at the beginning of the C--O shell merger ($\rm \Delta t_{coll} = 6.66\times10^{3}\ s$, central panel); and at the end of the C--O shell merger ($\rm \Delta t_{coll} = 5.72\times10^{2}\ s$, lower panel). After silicon is exhausted in the stellar core, the core begins its final contraction leading up to core collapse. This contraction lasts $\rm \sim 2\ days$ and raises the temperature at the base of the convective O burning shell from $\rm 2.58$ to $\rm 2.85\ GK$, allowing the convective zone to extend outward into the surrounding Ne, C and O rich layers (upper panel of \figurename~\ref{fig:abu}). While the bottom of the shell is receding in radius due to the contraction of the stellar core, the layers above expand due to the efficient convection. Once the outer boundary of the convective zone starts mixing in some new Ne, C, and O, a convective-reactive event is triggered (central panel of \figurename~\ref{fig:abu}). The ingested nuclei provide fresh fuel, which sustains further expansion of the convective zone and promotes additional ingestion of material from the surrounding layers. The C--O shell merger freezes once an active Si burning shell forms outside the Fe core, which accelerates the timescale of the final evolutionary stages of the massive star (lower panel of \figurename~\ref{fig:abu}).

    During a C--O shell merger, the nuclear energy that sustains the convective-reactive process is concentrated in a narrow region near the base of the shell. Within this region, the abundances of the O-burning products remain relatively constant, as their enhanced production is balanced by the outward expansion of the convective zone (see \tablename~\ref{tab:oburn}). This expansion helps diluting the newly synthesized material into layers where it is no longer processed, while also supplying fresh fuel to the burning region. As a result, O burning products end up being largely overproduced relative to their initial abundance in the entire shell.
    
    Compared to classical O shell burning, the main difference in a C--O shell merger lies in the much larger production of light particles ($p$, $n$, $\alpha$). The excess of these light particles is generated by: (1) the more efficient O burning itself (with branching ratios at these temperatures of roughly 66$\%$ for $p$ emission, 21$\%$ for $\alpha$, and 13$\%$ for $n$); Ne photodissociation (emission of $\alpha$ particles); C burning (about 41$\%$ for $p$ emission,  51$\%$ for $\alpha$, and 8$\%$ for $n$); (4) and hybrid C-O fusion (approximately 54$\%$ for $p$ emission, 36$\%$ for $\alpha$, and 10$\%$ for $n$). Altogether, these processes lead to a large enhancement in the number of protons (around 60 times more than in the case of O shell burning without C and Ne ingestion), a strong increase in $\alpha$ particles (about 30 times higher), and a significant rise in neutron production (also about 30 times higher). The higher $n$ abundance (of the order of $\sim2\times10^{-14}$ in mass fraction), combined with the typical density at the bottom of the O shell, results in a neutron density of about $n_{n}\sim 2\times 10^{16}\ cm^{-3}$. 
    
    The main neutron sources are \isotope[16]{O}(\isotope[16]{O},$n$)\isotope[31]{S} and \isotope[16]{O}(\isotope[12]{C},$n$)\isotope[27]{Si}. While we do not observe in this model a significant production of heavy elements through neutron captures, the availability of neutrons is crucial to increase the production of some short-lived radioactive nuclei (such as \isotope[36]{Cl}, \isotope[40]{K}, and \isotope[41]{Ca}), which signature can be measured in the Early Solar System material \citep[e.g.,][]{lugaro:18}. Beyond iron, these neutrons could also boost the production of some isotopes traditionally considered $r$-only, such as \isotope[70]{Zn}, \isotope[76]{Ge}, and \isotope[80]{Se}. This aspect can not be studied in detail here due to the low metallicity of our case of study, but it will need to be explored in detail for stellar models with C--O shell mergers with initial  metallicity closer to solar \citep[e.g.,][]{ritter:18a, roberti:23}. 
    
    The large number of protons plays a key role in the synthesis of odd-Z elements and in activating several nuclear processes that can significantly alter the energy balance within the evolving convective shell. 
   
\section{The SPAr burning} \label{sec:spar}

    During O burning, the O fusion is not the only dominant process, but it is accompanied by a large number of charged particle reactions such as \isotope[27]{Al}($p$,$\gamma$)\isotope[28]{Si}, \ptosi, and \isotope[28]{Si}($\alpha$,$\gamma$)\isotope[32]{S} \citep{thielemann:85,chieffi:98}. These charged particles are primarily produced by the \isotope[16]{O}+\isotope[16]{O} reactions and subsequently captured by the products of Ne burning (\isotope[24]{Mg}, \isotope[27]{Al}) as well as by products of O burning itself (\isotope[31]{P}, \isotope[28]{Si}). This results in a final chemical composition dominated by \isotope[28,30]{Si}, \isotope[32,34]{S}, and \isotope[38]{Ar}. The high temperatures and densities of O burning favour also the activation of efficient electron captures, $\beta$ decays, and photodisintegrations. In particular, these conditions allows the capture reactions to begin approaching an equilibrium between their forward and reverse channels. As described by \cite{chieffi:98}, the degree of equilibrium of a reaction $a+b\to c+d$ can be estimated by the quantity $\rm \varphi = \frac{|f_{ab}-f_{cd}|}{max(f_{ab},f_{cd})}$, where $f_{ab}$ and $f_{cd}$ are the fluxes\footnote{$\rm f_{ab} = Y_aY_b \rho\ N_A\left \langle \sigma v\right \rangle_{ab}$ (and $\rm f_{a\gamma} = Y_a N_A\lambda_{a\gamma}$ in the case of a photodisintegration), with $\rm Y_i=X_i/A_i$ where $\rm X_i$ and $\rm A_i$ are the mass fraction and the atomic weight of the species $i$, respectively.} of the forward and reverse reactions, respectively. A value of $\varphi \approx 0$ indicates that the process is close to equilibrium, whereas $\varphi \approx 1$ suggests it is far from equilibrium. 
    It is also worth noting that during O burning, some reactions initially dominated by the forward channel may reverse, becoming dominated by the reverse reaction as they approach small values of $\varphi$.

    \begin{figure}[!t]
        \centering
        \includegraphics[width=\linewidth]{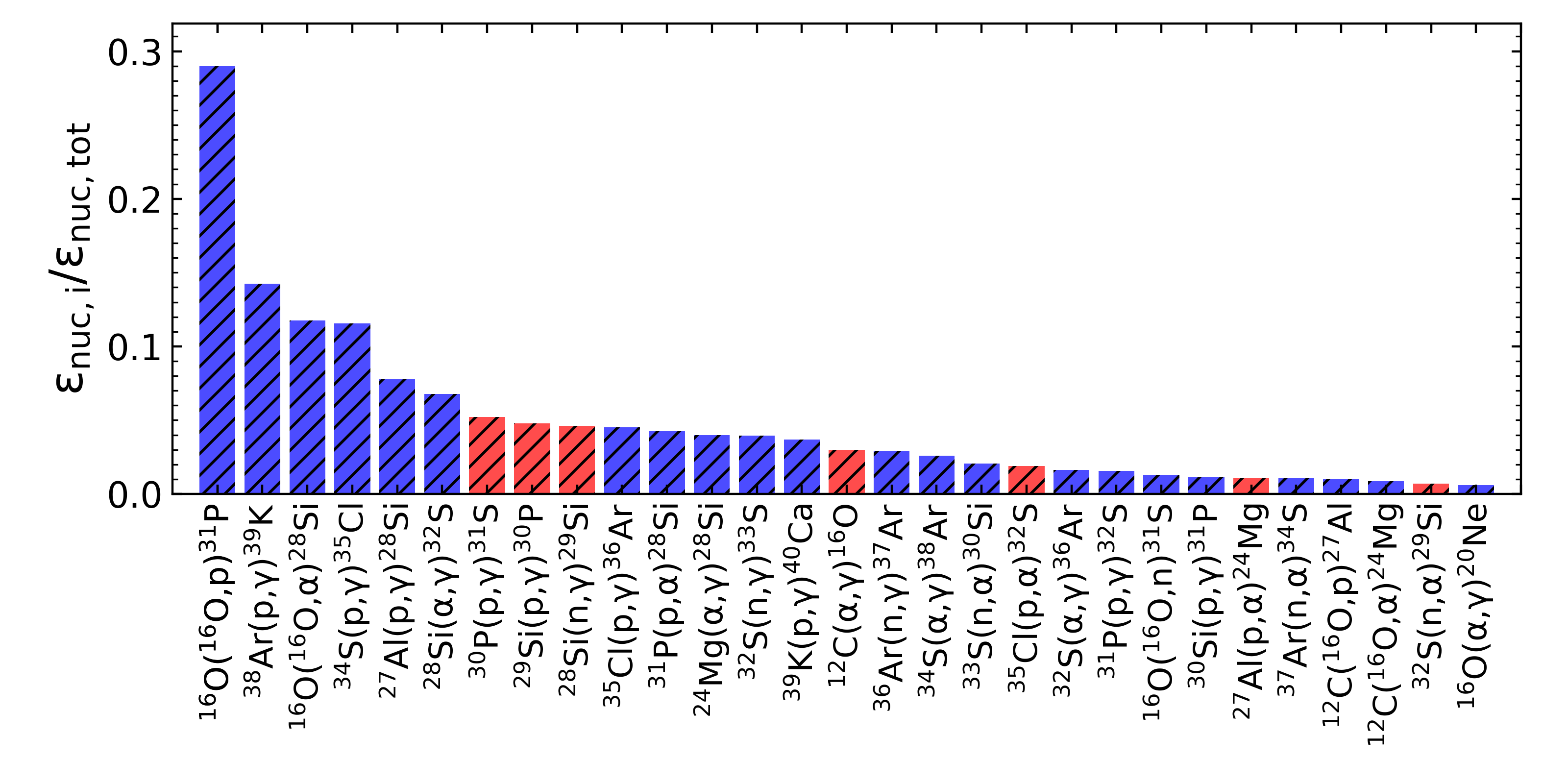}
        \includegraphics[width=\linewidth]{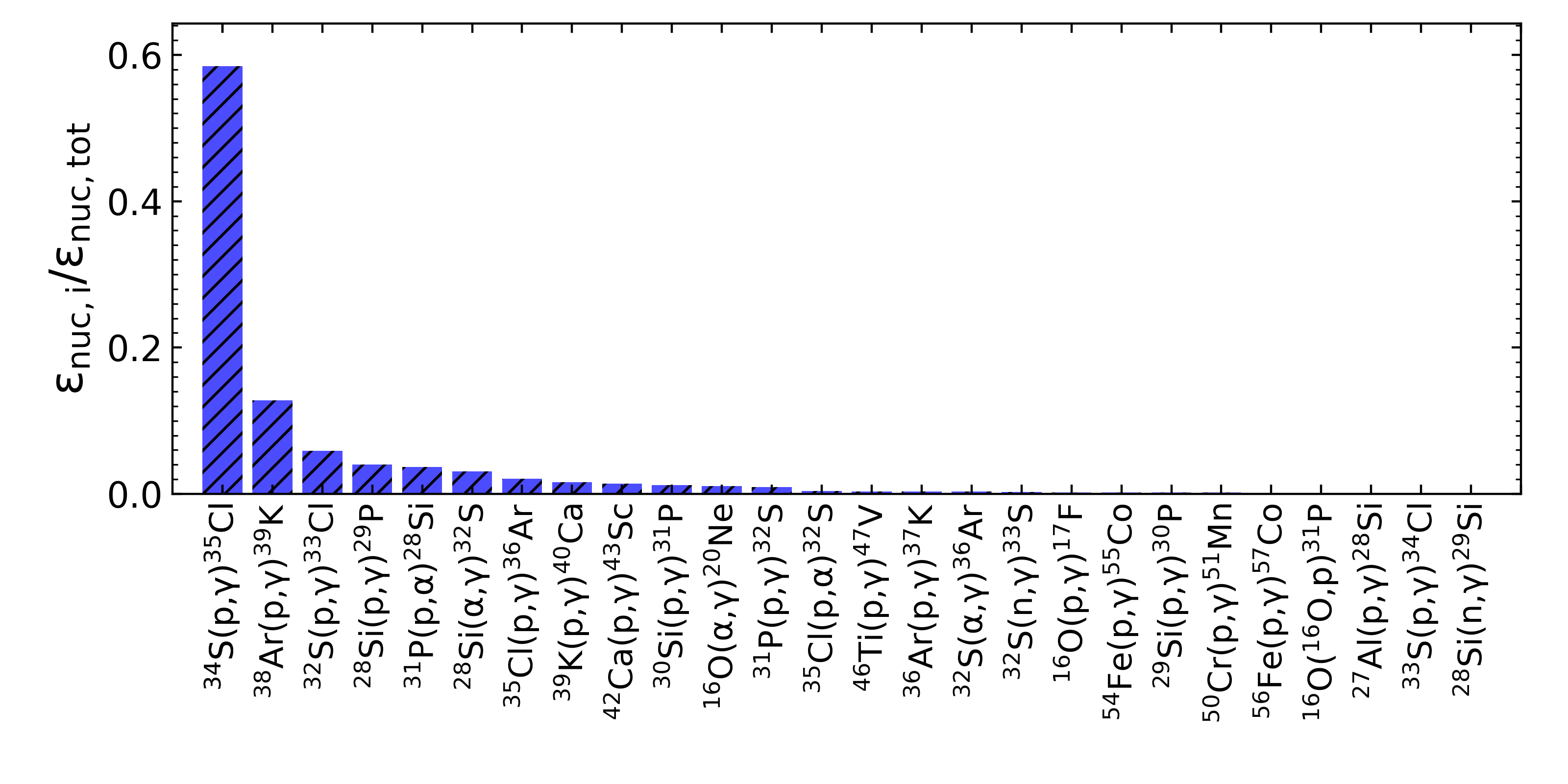}
        \includegraphics[width=\linewidth]{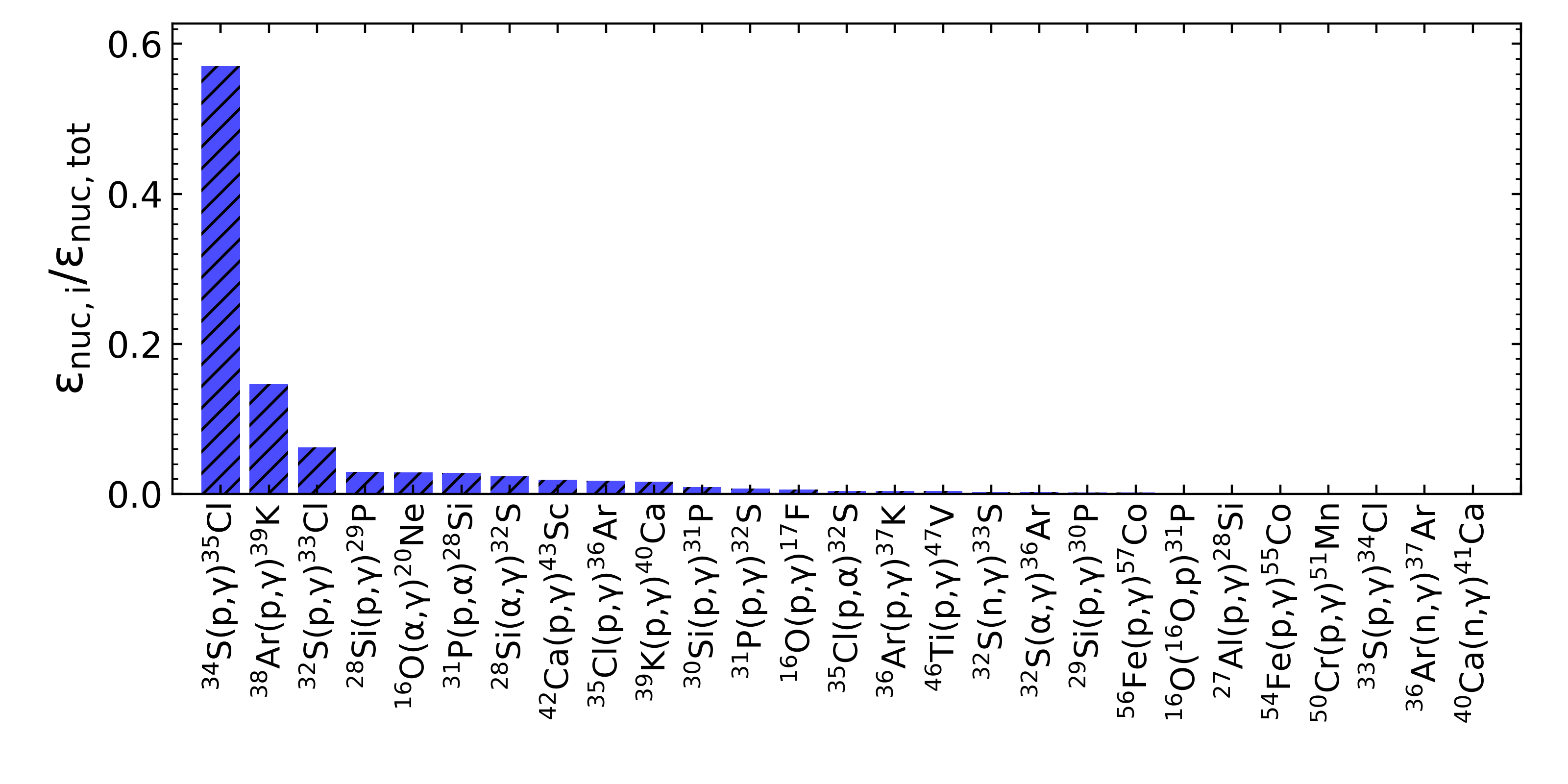}
        \caption{The relative contribution of individual effective nuclear reactions (forward minus reverse reaction, see text) to the total nuclear energy production at the bottom of the O-burning shell during O shell burning (upper panel), at the beginning of the C--O shell merger (central panel), and at the end of the C--O shell merger (lower panel). The blue bars represent processes that are dominated by the forward reaction and have a positive energy contribution, while red bars represent instead processes that are dominated by the reverse reaction and have a negative energy contribution. The three panels include all the reactions having an absolute $\varepsilon/\varepsilon_{tot}>10^{-3}$ (See also \tablename~\ref{tab:eratio1}-\ref{tab:eratio2}-\ref{tab:eratio3}).}
        \label{fig:eratio}
    \end{figure}
    
    The net energy produced by a nuclear reaction per unit time and mass ($\varepsilon_{nuc}$) can be estimated as the product of the Q-value of the reaction (the energy released, calculated from the mass difference between reactants and products) and the effective reaction flux, defined as the difference between the fluxes of the forward and reverse reactions. Summing over all reactions provides the total $\varepsilon_{nuc}$ for a given stellar zone, and integrating this quantity over the mass of the convective shell gives the total nuclear energy generated (or absorbed, if negative) per second within the shell. The zones at the highest temperatures, specifically those at the bottom of the O shell, contribute the most to the total energy produced per second (see the dotted-dashed line in central and lower panels of \figurename~\ref{fig:abu}, representing the cumulative $\varepsilon_{nuc}$). Therefore, they are representative of the nuclear processes that dominate energy generation within the shell. \figurename~\ref{fig:eratio} shows the relative contribution of individual nuclear reactions to the total nuclear energy production at the bottom of the O-burning shell, sorted in decreasing order of their absolute value (see also \tablename~\ref{tab:eratio1}-\ref{tab:eratio2}-\ref{tab:eratio3}).
    
    The energy generation and the overall reaction balance are significantly influenced by the much higher abundance of light particles present during a C--O shell merger. In particular, the large production of $p$ drives the radiative capture reactions ($p,\gamma$) out of equilibrium with their photodisintegration counterparts ($\gamma,p$), causing the capture channel to dominate.
    For example, in the case of the \stocl\ reaction, at the bottom of the shell $\varphi$ increases from $\varphi = 1.82\times10^{-3}$ in O shell burning, to $\varphi = 0.251$ and $\varphi = 0.469$ at the beginning and at the end of the merger, respectively. As a result, proton captures on S, P, and Ar (specifically, \stocll, \stocl, \ptosi, and \artok) become (1) the most efficient reactions and (2) the dominant contributors to energy generation throughout the entire duration of the merger, as soon as the abundance of protons exceeds a few times $10^{-9}$. Meanwhile, the contributions of O fusion to the total energy budget at the bottom of the C--O shell drops to around 0.13-0.14\% (see central and lower panels of \figurename~\ref{fig:eratio}). 
    The energy generation due to these proton-capture reactions also dominates the entire integral over the mass, and therefore it guides the temporal evolution of the convective-reactive event. We call this simultaneous activation of nuclear reactions due to proton captures the SPAr (standing for S, P, and Ar) burning. For the model considered in our analysis, the SPAr burning reactions generate about 400 times more energy than the typical \isotope[16]{O}--\isotope[16]{O} fusion channels and $\sim330$ times more energy than the sum of C, O, and C+O fusion channels (central and lower panel of \figurename~\ref{fig:eratio}). As we mentioned in Sect.~\ref{sec:model}, the stellar model by \cite{roberti:24} considered for this analysis shows really similar properties compared to the other 1D models experiencing C--O shell mergers. Therefore, the SPAr burning is a general feature of C--O shell mergers, and not a result depending on specific details of the stellar model. We expect its energy boost to also be present in 3D hydrodynamics simulations of C-–O shell mergers, because of the generality of the thermodynamical conditions driven by the same initial conditions of O fusion.
 
    The SPAr burning mainly depends on two ingredients: the protons, that control the efficiency of the forward proton capture reactions \stocll, \stocl, \artok, and \ptosi, and the temperature, that instead controls the efficiency of the photodisintegrations \isotope[33]{Cl}($\gamma$,$p$)\isotope[32]{S}, \isotope[35]{Cl}($\gamma$,$p$)\isotope[34]{S}, and \isotope[39]{K}($\gamma$,$p$)\isotope[38]{Ar}. Interestingly, the \ptosi\ reaction ranks among the most efficient reactions in a merger, exhibiting one of the highest effective fluxes (see \ref{tab:eratio2}-\ref{tab:eratio3}). Its Q-value, however, is lower than that of other radiative capture reactions (e.g., \isotope[28]{Si}($p$,$\gamma$)\isotope[33]{P}), making it a key regulator of the balance between $p$ and $\alpha$ particles rather than a dominant energy source. Reactants and products have instead a marginal contribution on the SPAr burning efficiency, as their variation is always contained between a factor of 2-3 (see \tablename~\ref{tab:oburn}). Note that the SPAr reactions are already well known to be relevant during O-burning, particularly as the amount of \isotope[16]{O} decreases in the convective shell \citep[see, e.g.,][]{thielemann:85,limongi:08}. The key difference in a shell undergoing SPAr burning is that the energy generated by SPAr reactions is orders of magnitude stronger than that from O-fusion reactions, due to the reasons discussed above.

\section{Discussion and conclusions} \label{sec:discussion}

    In this paper, we demonstrated that the light particles released during a C--O shell merger (in particular protons), not only impact the nucleosynthesis, but also play a key role in the energy balance of the evolving convective shell. We identified four crucial nuclear reactions that dominate energy generation once the proton abundance becomes sufficiently high: \stocll, \stocl, \ptosi, and \artok, that we referred to as the SPAr burning. The nuclear energy released by the SPAr reactions is 400 times greater than that produced by classical O-burning shell conditions. This result is crucial, as neglecting their contribution (often done to simplify and speed up stellar calculations) can lead to inaccurate modeling of the final evolutionary stages of massive stars, potentially altering their CCSN outcome \citep[we refer to][for a complete discussion on how the network size can affect the pre-supernova structure]{farmer:16}.

    C--O shell mergers are a convective-reactive events, i.e., the ingestion of fresh fuel at the outer boundary of the convective zone promotes further expansion of the shell and additional fuel ingestion. Therefore, simulating this occurrence requires that the equations governing the evolution of the chemical composition to be (1) coupled with the stellar structure equations and mixing processes, and (2) include an extended nuclear network that captures the energy feedback from light particle reactions such as those in the SPAr process. Our analysis suggests that the additional energy released by SPAr burning increases the nuclear luminosity, which in turn steepens the radiative gradient. As a result, we expect the convective merged zone to grow, at least more rapidly than in models that do not include these reactions. Of course, the actual expansion of the convective zone also depends on other factors, such as neutrino losses and the treatment of convective boundary mixing \cite[e.g.,][]{rizzuti:23,rizzuti:24,brinkman:24}. Therefore, this expectation needs to be verified through full stellar evolution calculations. Qualitatively, we also expect that 3D models would exhibit a similar behavior, and that the use of boosting factors on the driving luminosity \citep[as discussed by][and references therein]{rizzuti:22} could partially compensate for the absence of SPAr reactions. Nevertheless, this should be confirmed with appropriate hydrodynamics simulations.

    Uncertainties in reaction rates can significantly impact stellar evolution during the advanced burning stages, as shown by, e.g.,  \cite{fields:18}. Adopting the most up-to-date and accurate rate estimates is therefore essential for a more accurate exploration of late stellar evolution.
    In this work, we adopt the stellar model from \citet{roberti:24}, using SPAr reaction rates from Reaclib \citep{reaclib} and Starlib \citep{starlib} nuclear databases. Specifically, in the case of the \stocll\ reaction experimental data from \citet{iliadis:10} was adopted. For the \stocl\ reaction, a theoretical rate from \citet{reaclib} was used, although the current version of Reaclib is based on experimental data from \citet{setoodehnia:19}. The \stocl\ reaction was recently measured in inverse kinematics by the DRAGON collaboration \citep{lovely:21}, though it has so far been mostly explored in the context of classical nova nucleosynthesis to explain isotopic anomalies in pre-solar grains \citep{ward:25}. The \ptosi\ rate is based on experimental data from \citet{iliadis:10}, while the \artok\ rate remains theoretical and is also taken from \citet{reaclib}. For \stocll, \ptosi\ and \artok, no updated experimental data have been published since their inclusion in Starlib. The only update to \ptosi\ in Reaclib is a revision adopting $~26\%$ of the recommended median rate from \citet{iliadis:10}. Incorporating the most recent experimental results into future simulations will be essential to assess uncertainties and better constrain model predictions. 

    Finally, the conditions of temperature and density in a C--O shell merger allows the activation of efficient weak reactions. In the stellar model, the most efficient weak reactions are the $\beta^+$ decays of \isotope[27]{Si}, \isotope[30]{P}, \isotope[31]{S}, \isotope[34]{Cl}, \isotope[37]{Ar}, \isotope[38]{K} and the electron captures on \isotope[32]{S} and \isotope[35]{Cl}. Most of the unstable nuclei are adjacent to SPAr products, together with the two products of the neutron sources \isotope[16]{O}(\isotope[16]{O},$n$)\isotope[31]{S} and \isotope[16]{O}(\isotope[12]{C},$n$)\isotope[27]{Si}. For these isotopes, the weak interactions adopted here are from \cite{fuller:82}. The upcoming PANDORA project \citep[e.g.,][]{mascali:22} will make it possible to measure weak interaction rates (as well as opacities) directly under plasma conditions, and the C–O shell merger may represent another promising application for such investigations.

\begin{acknowledgements}
We thank the support from the NKFI via K-project 138031 and the Lend\"ulet Program LP2023-10 of the Hungarian Academy of Sciences. LR and MP acknowledge the support to NuGrid from JINA-CEE (NSF Grant PHY-1430152) and STFC (through the University of Hull’s Consolidated Grant ST/R000840/1), and ongoing access to {\tt viper}, the University of Hull High Performance Computing Facility. LR acknowledges the support from the ChETEC-INFRA -- Transnational Access Projects 22102724-ST and 23103142-ST and the PRIN URKA Grant Number \verb |prin_2022rjlwhn|. This work was supported by the European Union’s Horizon 2020 research and innovation programme (ChETEC-INFRA -- Project no. 101008324), and the IReNA network supported by US NSF AccelNet (Grant No. OISE-1927130). MP also acknowledges the support
from  the ERC Synergy Grant Programme (Geoastronomy, grant agreement number 101166936, Germany) and the ERC Consolidator Grant funding scheme (Project RADIOSTAR, G.A. n. 724560, Hungary). This work benefited from
interactions and workshops co-organized by The Center for Nuclear astrophysics Across Messengers (CeNAM) which is supported by the U.S. Department of Energy, Office of Science, Office of Nuclear Physics, under Award Number DE-SC0023128. We thank the two anonymous referees for their valuable contribution.
\end{acknowledgements}

\appendix

\section{The stellar model}\label{app:1}

The model adopted in this work is part of a larger set calculated with the FRANEC code \citep{CL13,LC18}, with a nuclear network that includes 524 isotopes and more than 3000 reactions, fully coupled to the physical evolution of the star (the list of nuclear species included in the network, up to Zn, is provided in Table~\ref{tab:net}). The network is an extension of that adopted in \cite{LC18}, designed to explore neutron-capture nucleosynthesis in fast-rotating stars at extremely low metallicity. The choice of this network is particularly suited to follow the energy generation in the most advanced stages of the evolution. In particular, the set of nuclear species involved in the energy generation is fully consistent with those adopted in other works addressing this topic, such as the 127-isotope network of \cite{farmer:16}, the 204-isotope networks of \cite{brinkman:19,brinkman:21,brinkman:23}, and the 112-isotope network of \cite{limongi:24}. A detailed discussion of the network selection, the complete list of isotopes, as well as the Kippenhahn diagram of the model presented here, can be found in Section 2 and in Figure 10 of \cite{roberti:24}. 

    \begin{table}
        \caption{The nuclear network from n to Zn (see text). $\rm A_{min}$ and $\rm A_{max}$ are the minimum and maximum atomic weight of each element.}
        \label{tab:net}
	    \centering
        \begin{tabular}{lccc|lccc}
            \hline\hline
            Element & Z & $\rm A_{min}$ & $\rm A_{max}$ & Element & Z & $\rm A_{min}$ & $\rm A_{max}$\\
            \hline
            n  & 0  & 1   & 1  & S  & 16 & 31  & 37 \\ 
            H  & 1  & 1   & 3  & Cl & 17 & 33  & 38 \\ 
            He & 2  & 3   & 4  & Ar & 18 & 35  & 41 \\ 
            Li & 3  & 6   & 7  & K  & 19 & 37  & 44 \\ 
            Be & 4  & 7   & 10 & Ca & 20 & 39  & 49 \\ 
            B  & 5  & 10  & 11 & Sc & 21 & 41  & 49 \\ 
            C  & 6  & 12  & 14 & Ti & 22 & 44  & 51 \\ 
            N  & 7  & 13  & 16 & V  & 23 & 45  & 52 \\ 
            O  & 8  & 14  & 19 & Cr & 24 & 47  & 55 \\ 
            F  & 9  & 17  & 20 & Mn & 25 & 49  & 57 \\ 
            Ne & 10 & 19  & 23 & Fe & 26 & 51  & 61 \\ 
            Na & 11 & 21  & 24 & Co & 27 & 53  & 62 \\ 
            Mg & 12 & 23  & 27 & Ni & 28 & 55  & 65 \\ 
            Al & 13 & 25  & 28 & Cu & 29 & 57  & 66 \\ 
            Si & 14 & 27  & 32 & Zn & 30 & 60  & 71 \\ 
            P  & 15 & 29  & 34 &    &    &     &    \\
            \hline
        \end{tabular}
    \end{table}
    
Note that the abundances of typical O burning products such as \isotope[28]{Si}, \isotope[31]{P}, \isotope[32, 34]{S}, \isotope[33, 35]{Cl}, \isotope[38]{Ar}, and \isotope[39]{K}, are entirely independent from both initial stellar metallicity and rotation. This is because the nucleosynthesis of these species is predominantly primary, driven directly by the O fusion processes and the previous burning stages, with a negligible contribution from the pristine chemical composition. Moreover, rotational instabilities act on much longer timescales than that of the O-burning phase itself, therefore rotationally induced mixing is no longer effective at this stage. Additionally, the key conditions (chemical composition, temperature, and density in the O shell) in this model are comparable to those in other models with initial masses between roughly 12 and 30 \msun\ across different sets \citep{LC18,ritter:18,brinkman:21,RLC24,limongi:25}. Therefore, the adoption of this model is suitable to discuss the general properties of the O burning in shell. We also note that structural differences (e.g., the location of the C shell) do not lead to quantitative differences in the nuclear reactions occurring at the base of the C–O shell merging zone. Notice also that C--O shell mergers can be obtained at any metallicities and using different stellar codes for the simulations \citep[e.g.,][]{roberti:25}. For instance, massive star models developing C--O shell mergers are obtained at solar metallicity also using MESA \citep[][]{ritter:18, ritter:18a, roberti:23} and KEPLER \citep[][]{rauscher:02}. 

\begin{figure}[!t]
       \centering
       \includegraphics[width=\linewidth]{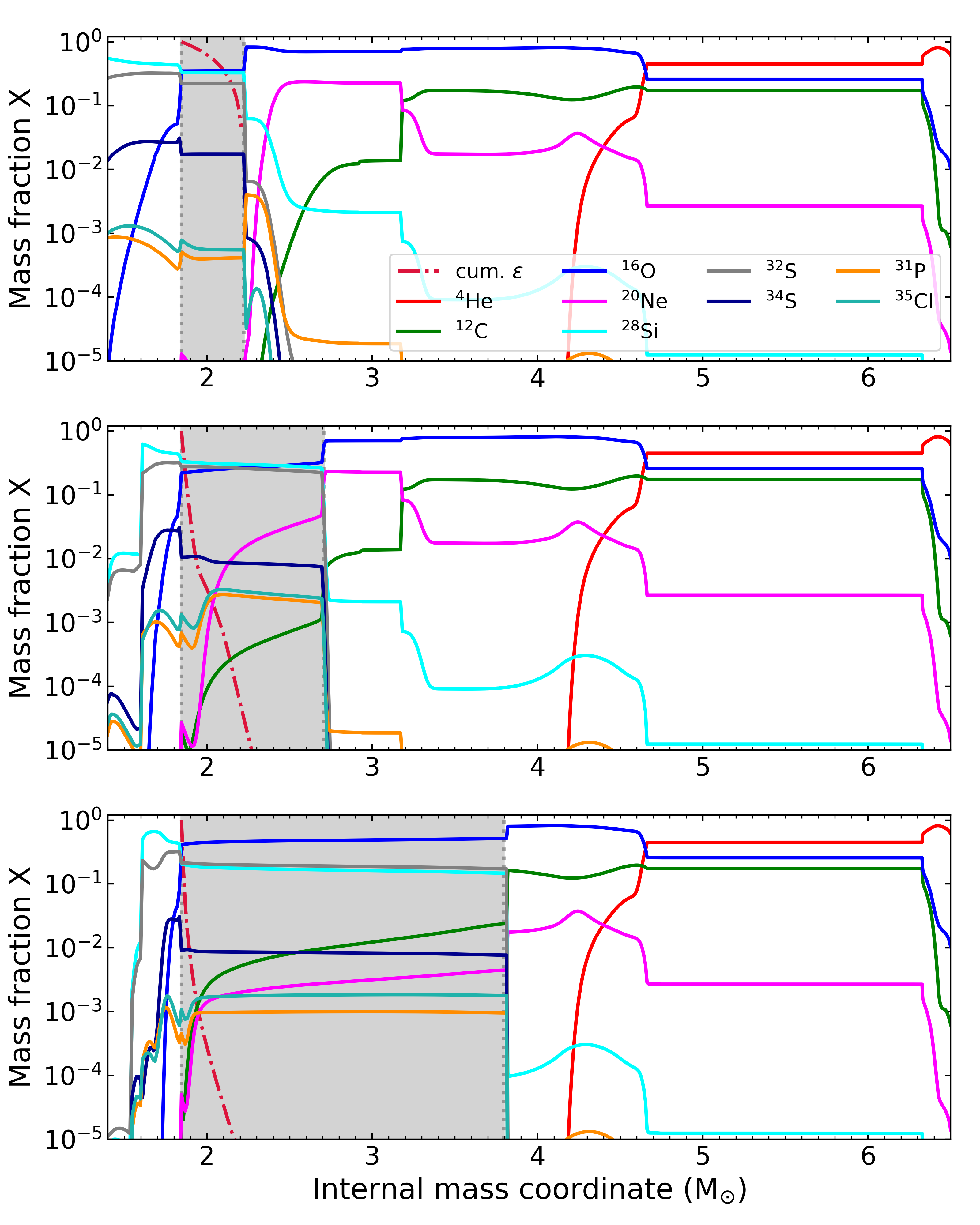}
       \caption{Abundances in mass fraction as a function of the internal mass coordinate for key nuclear species (see legend). The grey shaded area represents the O convective shell $\rm 3.47\times10^{5}\ s$ (upper panel), $\rm 6.66\times10^{3}\ s$ (central panel) and $\rm 5.72\times10^{2}\ s$ (lower panel) before the core-collapse.
       Neutron and proton abundances (not shown in the figure) with other relevant species are provided at these different stages in Table \ref{tab:oburn}, at the base of the grey zone.
       The dotted-dashed red line shows the cumulative integral of the energy per second generated by the nuclear reactions from the outer edge to the bottom of the O shell normalized to its maximum.}
       \label{fig:abu}
   \end{figure}

    In the following, we present the supplementary material relative to the chemical composition (\figurename~\ref{fig:abu}, \tablename~\ref{tab:oburn}) and to the most energetic nuclear reactions (\tablename~\ref{tab:eratio1}-\ref{tab:eratio2}-\ref{tab:eratio3}) at the bottom of the O-shell, for the three different timesteps reported in upper, central, and lower panels of \figurename~\ref{fig:eratio}.
            
    \begin{table}
        \caption{Stellar conditions and chemical abundances at the bottom of the O-burning shell (see text).}
        \label{tab:oburn}
	    \centering
        \begin{tabular}{cccc}
            \hline\hline
            & O-    & C-O              & C-O               \\
            & shell & shell merger (i) & shell merger (f)  \\
            \hline
            $\Delta t_{coll}$ (s)                 & 3.47e+04 & 6.66e+03 & 5.72e+02  \\
            T (GK)                                & 2.58     & 2.85     & 2.78      \\ 
            $\rho$ ($\rm 10^6\ g\ cm^{-3}$)  & 2.14     & 2.53     & 2.09      \\ 
            \hline
            $p$              & 1.54e-10 & 9.39e-09 & 7.21e-09 \\     
            $\alpha$         & 8.35e-09 & 2.86e-07 & 2.75e-07 \\        
            $n$              & 8.24e-16 & 2.32e-14 & 1.87e-14 \\
            \isotope[12]{C}  & 2.90e-06 & 2.01e-05 & 2.29e-05 \\    
            \isotope[16]{O}  & 3.48e-01 & 2.20e-01 & 4.11e-01 \\   
            \isotope[24]{Mg} & 1.57e-03 & 2.47e-04 & 4.66e-04 \\  
            \isotope[27]{Al} & 2.88e-05 & 5.65e-06 & 1.13e-05 \\    
            \isotope[28]{Si} & 3.29e-01 & 3.30e-01 & 2.04e-01 \\    
            \isotope[31]{P}  & 5.09e-04 & 6.82e-04 & 4.54e-04 \\   
            \isotope[32]{S}  & 2.22e-01 & 2.75e-01 & 2.13e-01 \\    
            \isotope[34]{S}  & 1.74e-02 & 1.06e-02 & 9.15e-03 \\  
            \isotope[33]{Cl} & 8.89e-11 & 2.00e-09 & 9.79e-10 \\   
            \isotope[35]{Cl} & 7.83e-04 & 1.33e-03 & 1.08e-03 \\    
            \isotope[38]{Ar} & 2.39e-02 & 2.32e-02 & 2.38e-02 \\     
            \isotope[39]{K}  & 1.11e-03 & 3.02e-03 & 2.90e-03 \\  
            \hline
        \end{tabular}
    \end{table}

    \begin{table*}
        \caption{List of the reactions that contribute the most to the energy generation at the bottom of the O-shell relative to the upper panels of \figurename~\ref{fig:abu}-\ref{fig:eratio}.}
        \label{tab:eratio1}
	   \centering
        \begin{tabular}{lccccc}
            \hline\hline
            Reaction & Q (MeV) & $\rm f_{ab}-f_{cd}$ & $\varepsilon_{i}$ & $\varepsilon_{i}$/$\varepsilon_{tot}$ & $\varphi$ \\
            \hline
                      $^{16}$O($^{16}$O,p)$^{31}$P &  7.68 &  4.17e+18 &  5.13e+13 & 2.90e-01 & 1.00e+00 \\
                 $^{38}$Ar(p,$\rm \gamma$)$^{39}$K &  6.38 &  2.47e+18 &  2.52e+13 & 1.42e-01 & 1.65e-02 \\
          $^{16}$O($^{16}$O,$\rm \alpha$)$^{28}$Si &  9.59 &  1.36e+18 &  2.08e+13 & 1.18e-01 & 1.00e+00 \\
                 $^{34}$S(p,$\rm \gamma$)$^{35}$Cl &  6.37 &  2.01e+18 &  2.04e+13 & 1.16e-01 & 1.82e-03 \\
                $^{27}$Al(p,$\rm \gamma$)$^{28}$Si & 11.58 &  7.41e+17 &  1.37e+13 & 7.77e-02 & 1.00e+00 \\
      $^{28}$Si($\rm \alpha$,$\rm \gamma$)$^{32}$S &  6.95 &  1.08e+18 &  1.20e+13 & 6.77e-02 & 3.15e-02 \\
                  $^{30}$P(p,$\rm \gamma$)$^{31}$S &  6.13 & -9.39e+17 & -9.22e+12 & 5.21e-02 & 9.97e-01 \\
                 $^{29}$Si(p,$\rm \gamma$)$^{30}$P &  5.59 & -9.43e+17 & -8.44e+12 & 4.77e-02 & 9.69e-02 \\
                $^{28}$Si(n,$\rm \gamma$)$^{29}$Si &  8.47 & -6.03e+17 & -8.18e+12 & 4.63e-02 & 2.28e-01 \\
                $^{35}$Cl(p,$\rm \gamma$)$^{36}$Ar &  8.51 &  5.88e+17 &  8.01e+12 & 4.53e-02 & 7.21e-02 \\
                 $^{31}$P(p,$\rm \alpha$)$^{28}$Si &  1.92 &  2.46e+18 &  7.53e+12 & 4.26e-02 & 1.00e-02 \\
     $^{24}$Mg($\rm \alpha$,$\rm \gamma$)$^{28}$Si &  9.98 &  4.43e+17 &  7.08e+12 & 4.00e-02 & 1.00e+00 \\
                  $^{32}$S(n,$\rm \gamma$)$^{33}$S &  8.64 &  5.03e+17 &  6.96e+12 & 3.94e-02 & 1.29e-01 \\
                 $^{39}$K(p,$\rm \gamma$)$^{40}$Ca &  8.33 &  4.87e+17 &  6.49e+12 & 3.67e-02 & 1.23e-01 \\
       $^{12}$C($\rm \alpha$,$\rm \gamma$)$^{16}$O &  7.16 & -4.63e+17 & -5.31e+12 & 3.00e-02 & 1.00e+00 \\
                $^{36}$Ar(n,$\rm \gamma$)$^{37}$Ar &  8.79 &  3.69e+17 &  5.19e+12 & 2.93e-02 & 4.68e-01 \\
      $^{34}$S($\rm \alpha$,$\rm \gamma$)$^{38}$Ar &  7.21 &  3.97e+17 &  4.58e+12 & 2.59e-02 & 3.54e-01 \\
                 $^{33}$S(n,$\rm \alpha$)$^{30}$Si &  3.49 &  6.52e+17 &  3.64e+12 & 2.06e-02 & 8.64e-01 \\
                 $^{35}$Cl(p,$\rm \alpha$)$^{32}$S &  1.87 & -1.12e+18 & -3.36e+12 & 1.90e-02 & 4.51e-02 \\
      $^{32}$S($\rm \alpha$,$\rm \gamma$)$^{36}$Ar &  6.64 &  2.73e+17 &  2.90e+12 & 1.64e-02 & 1.14e-01 \\
                  $^{31}$P(p,$\rm \gamma$)$^{32}$S &  8.86 &  1.96e+17 &  2.79e+12 & 1.58e-02 & 4.12e-02 \\
                      $^{16}$O($^{16}$O,n)$^{31}$S &  1.50 &  9.47e+17 &  2.27e+12 & 1.29e-02 & 1.00e+00 \\
                 $^{30}$Si(p,$\rm \gamma$)$^{31}$P &  7.30 &  1.70e+17 &  1.98e+12 & 1.12e-02 & 1.05e-02 \\
                $^{27}$Al(p,$\rm \alpha$)$^{24}$Mg &  1.60 & -7.56e+17 & -1.94e+12 & 1.09e-02 & 7.56e-02 \\
                 $^{37}$Ar(n,$\rm \alpha$)$^{34}$S &  4.63 &  2.60e+17 &  1.92e+12 & 1.09e-02 & 7.74e-01 \\
                     $^{12}$C($^{16}$O,p)$^{27}$Al &  5.17 &  2.13e+17 &  1.76e+12 & 9.98e-03 & 1.00e+00 \\
          $^{12}$C($^{16}$O,$\rm \alpha$)$^{24}$Mg &  6.77 &  1.44e+17 &  1.56e+12 & 8.80e-03 & 1.00e+00 \\
                 $^{32}$S(n,$\rm \alpha$)$^{29}$Si &  1.53 & -5.04e+17 & -1.23e+12 & 6.95e-03 & 2.52e-01 \\
      $^{16}$O($\rm \alpha$,$\rm \gamma$)$^{20}$Ne &  4.73 &  1.40e+17 &  1.06e+12 & 6.01e-03 & 2.40e-03 \\
                 $^{39}$K(p,$\rm \alpha$)$^{36}$Ar &  1.29 & -4.25e+17 & -8.76e+11 & 4.95e-03 & 3.09e-01 \\
                $^{27}$Al($\rm \alpha$,p)$^{30}$Si &  2.37 &  1.93e+17 &  7.31e+11 & 4.13e-03 & 1.00e+00 \\
                $^{37}$Cl(p,$\rm \gamma$)$^{38}$Ar & 10.24 &  3.87e+16 &  6.34e+11 & 3.58e-03 & 7.18e-01 \\
                $^{40}$Ca(n,$\rm \gamma$)$^{41}$Ca &  8.36 &  4.33e+16 &  5.80e+11 & 3.28e-03 & 1.11e-01 \\
                 $^{37}$Cl(p,$\rm \alpha$)$^{34}$S &  3.03 &  1.09e+17 &  5.29e+11 & 2.99e-03 & 5.64e-01 \\
      $^{29}$Si($\rm \alpha$,$\rm \gamma$)$^{33}$S &  7.12 &  3.42e+16 &  3.90e+11 & 2.20e-03 & 3.49e-01 \\
                           $^{37}$Ar(n,p)$^{37}$Cl &  1.60 &  1.48e+17 &  3.78e+11 & 2.14e-03 & 4.82e-01 \\
                $^{41}$Ca(n,$\rm \alpha$)$^{38}$Ar &  5.22 &  4.20e+16 &  3.51e+11 & 1.98e-03 & 8.56e-01 \\
                  $^{33}$S(n,$\rm \gamma$)$^{34}$S & 11.42 &  1.89e+16 &  3.45e+11 & 1.95e-03 & 8.78e-01 \\
                 $^{32}$S(p,$\rm \gamma$)$^{33}$Cl &  2.28 &  9.34e+16 &  3.40e+11 & 1.92e-03 & 1.84e-04 \\
                $^{36}$Cl(p,$\rm \gamma$)$^{37}$Ar &  8.72 &  2.37e+16 &  3.30e+11 & 1.87e-03 & 4.44e-01 \\
     $^{20}$Ne($\rm \alpha$,$\rm \gamma$)$^{24}$Mg &  9.32 &  1.48e+16 &  2.21e+11 & 1.25e-03 & 9.97e-01 \\
      $^{33}$S($\rm \alpha$,$\rm \gamma$)$^{37}$Ar &  6.79 &  2.02e+16 &  2.20e+11 & 1.24e-03 & 4.59e-01 \\
      $^{30}$Si($\rm \alpha$,$\rm \gamma$)$^{34}$S &  7.92 &  1.41e+16 &  1.78e+11 & 1.01e-03 & 9.94e-02 \\
            \hline
        \end{tabular}
    \end{table*} 

    \begin{table*}
        \caption{List of the reactions that contribute the most to the energy generation at the bottom of the O-shell relative to the central panels of \figurename~\ref{fig:abu}-\ref{fig:eratio}.}
        \label{tab:eratio2}
	   \centering
        \begin{tabular}{lccccc}
            \hline\hline
            Reaction & Q (MeV) & $\rm f_{ab}-f_{cd}$ & $\varepsilon_{i}$ & $\varepsilon_{i}$/$\varepsilon_{tot}$ & $\varphi$ \\
            \hline
                 $^{34}$S(p,$\rm \gamma$)$^{35}$Cl &  6.37 & 1.26e+22 & 1.29e+17 & 5.84e-01 & 2.51e-01 \\
                 $^{38}$Ar(p,$\rm \gamma$)$^{39}$K &  6.38 & 2.76e+21 & 2.82e+16 & 1.28e-01 & 2.53e-01 \\
                 $^{32}$S(p,$\rm \gamma$)$^{33}$Cl &  2.28 & 3.54e+21 & 1.29e+16 & 5.86e-02 & 9.14e-02 \\
                 $^{28}$Si(p,$\rm \gamma$)$^{29}$P &  2.75 & 2.01e+21 & 8.86e+15 & 4.02e-02 & 1.12e-01 \\
                 $^{31}$P(p,$\rm \alpha$)$^{28}$Si &  1.92 & 2.63e+21 & 8.07e+15 & 3.67e-02 & 8.76e-02 \\
      $^{28}$Si($\rm \alpha$,$\rm \gamma$)$^{32}$S &  6.95 & 6.09e+20 & 6.77e+15 & 3.08e-02 & 2.70e-01 \\
                $^{35}$Cl(p,$\rm \gamma$)$^{36}$Ar &  8.51 & 3.31e+20 & 4.50e+15 & 2.04e-02 & 3.36e-01 \\
                 $^{39}$K(p,$\rm \gamma$)$^{40}$Ca &  8.33 & 2.68e+20 & 3.57e+15 & 1.62e-02 & 3.30e-01 \\
                $^{42}$Ca(p,$\rm \gamma$)$^{43}$Sc &  4.93 & 3.97e+20 & 3.13e+15 & 1.42e-02 & 1.96e-01 \\
                 $^{30}$Si(p,$\rm \gamma$)$^{31}$P &  7.30 & 2.30e+20 & 2.68e+15 & 1.22e-02 & 2.88e-01 \\
      $^{16}$O($\rm \alpha$,$\rm \gamma$)$^{20}$Ne &  4.73 & 3.09e+20 & 2.34e+15 & 1.06e-02 & 1.92e-01 \\
                  $^{31}$P(p,$\rm \gamma$)$^{32}$S &  8.86 & 1.49e+20 & 2.12e+15 & 9.62e-03 & 3.34e-01 \\
                 $^{35}$Cl(p,$\rm \alpha$)$^{32}$S &  1.87 & 3.05e+20 & 9.10e+14 & 4.14e-03 & 7.81e-02 \\
                 $^{46}$Ti(p,$\rm \gamma$)$^{47}$V &  5.17 & 9.53e+19 & 7.88e+14 & 3.58e-03 & 2.07e-01 \\
                 $^{36}$Ar(p,$\rm \gamma$)$^{37}$K &  1.86 & 2.33e+20 & 6.92e+14 & 3.14e-03 & 7.35e-02 \\
      $^{32}$S($\rm \alpha$,$\rm \gamma$)$^{36}$Ar &  6.64 & 6.30e+19 & 6.69e+14 & 3.04e-03 & 2.80e-01 \\
                  $^{32}$S(n,$\rm \gamma$)$^{33}$S &  8.64 & 4.62e+19 & 6.39e+14 & 2.90e-03 & 3.42e-01 \\
                  $^{16}$O(p,$\rm \gamma$)$^{17}$F &  0.60 & 4.55e+20 & 4.37e+14 & 1.98e-03 & 1.63e-02 \\
                $^{54}$Fe(p,$\rm \gamma$)$^{55}$Co &  5.06 & 5.38e+19 & 4.36e+14 & 1.98e-03 & 2.05e-01 \\
                 $^{29}$Si(p,$\rm \gamma$)$^{30}$P &  5.59 & 4.67e+19 & 4.18e+14 & 1.90e-03 & 1.98e-01 \\
                $^{50}$Cr(p,$\rm \gamma$)$^{51}$Mn &  5.27 & 4.42e+19 & 3.73e+14 & 1.69e-03 & 2.14e-01 \\
                $^{56}$Fe(p,$\rm \gamma$)$^{57}$Co &  6.03 & 2.99e+19 & 2.88e+14 & 1.31e-03 & 2.44e-01 \\
                      $^{16}$O($^{16}$O,p)$^{31}$P &  7.68 & 2.30e+19 & 2.83e+14 & 1.28e-03 & 1.00e+00 \\
                $^{27}$Al(p,$\rm \gamma$)$^{28}$Si & 11.58 & 1.35e+19 & 2.50e+14 & 1.13e-03 & 9.96e-01 \\
                 $^{33}$S(p,$\rm \gamma$)$^{34}$Cl &  5.14 & 2.92e+19 & 2.40e+14 & 1.09e-03 & 2.01e-01 \\
                $^{28}$Si(n,$\rm \gamma$)$^{29}$Si &  8.47 & 1.63e+19 & 2.21e+14 & 1.00e-03 & 2.89e-01 \\
            \hline
        \end{tabular}
    \end{table*} 

    \begin{table*}
        \caption{List of the reactions that contribute the most to the energy generation at the bottom of the O-shell relative to the lower panels of \figurename~\ref{fig:abu}-\ref{fig:eratio}.}
        \label{tab:eratio3}
	   \centering
        \begin{tabular}{lccccc}
            \hline\hline
            Reaction & Q (MeV) & $\rm f_{ab}-f_{cd}$ & $\varepsilon_{i}$ & $\varepsilon_{i}$/$\varepsilon_{tot}$ & $\varphi$ \\
            \hline
                 $^{34}$S(p,$\rm \gamma$)$^{35}$Cl &  6.37 & 8.97e+21 & 9.15e+16 & 5.70e-01 & 4.69e-01 \\
                 $^{38}$Ar(p,$\rm \gamma$)$^{39}$K &  6.38 & 2.30e+21 & 2.35e+16 & 1.46e-01 & 4.71e-01 \\
                 $^{32}$S(p,$\rm \gamma$)$^{33}$Cl &  2.28 & 2.72e+21 & 9.92e+15 & 6.18e-02 & 1.93e-01 \\
                 $^{28}$Si(p,$\rm \gamma$)$^{29}$P &  2.75 & 1.08e+21 & 4.73e+15 & 2.95e-02 & 2.31e-01 \\
      $^{16}$O($\rm \alpha$,$\rm \gamma$)$^{20}$Ne &  4.73 & 6.10e+20 & 4.62e+15 & 2.88e-02 & 3.73e-01 \\
                 $^{31}$P(p,$\rm \alpha$)$^{28}$Si &  1.92 & 1.48e+21 & 4.54e+15 & 2.83e-02 & 1.81e-01 \\
      $^{28}$Si($\rm \alpha$,$\rm \gamma$)$^{32}$S &  6.95 & 3.40e+20 & 3.78e+15 & 2.35e-02 & 5.08e-01 \\
                $^{42}$Ca(p,$\rm \gamma$)$^{43}$Sc &  4.93 & 3.90e+20 & 3.08e+15 & 1.92e-02 & 3.80e-01 \\
                $^{35}$Cl(p,$\rm \gamma$)$^{36}$Ar &  8.51 & 2.09e+20 & 2.85e+15 & 1.78e-02 & 5.88e-01 \\
                 $^{39}$K(p,$\rm \gamma$)$^{40}$Ca &  8.33 & 1.96e+20 & 2.61e+15 & 1.63e-02 & 5.80e-01 \\
                 $^{30}$Si(p,$\rm \gamma$)$^{31}$P &  7.30 & 1.30e+20 & 1.51e+15 & 9.44e-03 & 5.31e-01 \\
                  $^{31}$P(p,$\rm \gamma$)$^{32}$S &  8.86 & 8.03e+19 & 1.14e+15 & 7.10e-03 & 5.97e-01 \\
                  $^{16}$O(p,$\rm \gamma$)$^{17}$F &  0.60 & 9.38e+20 & 9.01e+14 & 5.62e-03 & 4.11e-02 \\
                 $^{35}$Cl(p,$\rm \alpha$)$^{32}$S &  1.87 & 2.09e+20 & 6.23e+14 & 3.88e-03 & 1.66e-01 \\
                 $^{36}$Ar(p,$\rm \gamma$)$^{37}$K &  1.86 & 2.05e+20 & 6.09e+14 & 3.79e-03 & 1.58e-01 \\
                 $^{46}$Ti(p,$\rm \gamma$)$^{47}$V &  5.17 & 7.04e+19 & 5.82e+14 & 3.63e-03 & 3.99e-01 \\
                  $^{32}$S(n,$\rm \gamma$)$^{33}$S &  8.64 & 3.23e+19 & 4.46e+14 & 2.78e-03 & 5.96e-01 \\
      $^{32}$S($\rm \alpha$,$\rm \gamma$)$^{36}$Ar &  6.64 & 3.87e+19 & 4.11e+14 & 2.56e-03 & 5.06e-01 \\
                 $^{29}$Si(p,$\rm \gamma$)$^{30}$P &  5.59 & 3.43e+19 & 3.07e+14 & 1.91e-03 & 3.73e-01 \\
                $^{56}$Fe(p,$\rm \gamma$)$^{57}$Co &  6.03 & 2.83e+19 & 2.73e+14 & 1.70e-03 & 4.56e-01 \\
                      $^{16}$O($^{16}$O,p)$^{31}$P &  7.68 & 1.82e+19 & 2.24e+14 & 1.40e-03 & 1.00e+00 \\
                $^{27}$Al(p,$\rm \gamma$)$^{28}$Si & 11.58 & 1.21e+19 & 2.24e+14 & 1.39e-03 & 1.00e+00 \\
                $^{54}$Fe(p,$\rm \gamma$)$^{55}$Co &  5.06 & 2.61e+19 & 2.12e+14 & 1.32e-03 & 3.95e-01 \\
                $^{50}$Cr(p,$\rm \gamma$)$^{51}$Mn &  5.27 & 2.43e+19 & 2.05e+14 & 1.28e-03 & 4.09e-01 \\
                 $^{33}$S(p,$\rm \gamma$)$^{34}$Cl &  5.14 & 2.41e+19 & 1.99e+14 & 1.24e-03 & 3.89e-01 \\
                $^{36}$Ar(n,$\rm \gamma$)$^{37}$Ar &  8.79 & 1.29e+19 & 1.81e+14 & 1.13e-03 & 6.65e-01 \\
                $^{40}$Ca(n,$\rm \gamma$)$^{41}$Ca &  8.36 & 1.20e+19 & 1.61e+14 & 1.00e-03 & 5.85e-01 \\
            \hline
        \end{tabular}
    \end{table*}

\end{document}